\documentclass[10pt]{article}
\usepackage[OE]{express_arxiv}
\usepackage{gensymb}
\usepackage{subfig}
\usepackage[title]{appendix}
\usepackage{amsmath}

\begin{document}
\title{Multi-reflection model of subwavelength grating diffraction based on simplified modal method }

\author{Changcheng Xiang,\authormark{1} Jun Wu,\authormark{2}, Changhe Zhou\authormark{1*} and Wei Jia,\authormark{1}}

\address{
\authormark{1}Institute of Photonics Technology, Jinan University, Guangdong, 510632, China\\
\authormark{2}Zhejiang University of Science and technology, No. 318, Hangzhou, Zhejiang, 310023, China\\}

\email{\authormark{*}chazhou@mail.shcnc.ac.cn} %% email address is required



\begin{abstract}
A multi-reflection model of grating diffraction based on the simplified modal method is proposed. Simulation results for a guided mode resonance Brewster grating using our method and rigorous coupled wave analysis are presented to verify our model. The solution  of our method is in good agreement with that of rigorous coupled wave analysis. Benefiting from its clear physical view, this model helps us to better understand the diffraction process inside subwavelength gratings. On the basis of the multi-reflection model, we developed a matrix Fabry--Perot (FP) resonance condition and a single-mode resonance condition to evaluate the resonance wavelength. These resonance conditions may be helpful for simplifying design of guided mode resonance (GMR) gratings.
\end{abstract}

 % REPLACE WITH CORRECT OCIS CODES FOR YOUR ARTICLE, MINIMUM OF TWO; Avoid using the OCIS codes for “General” or “General science” whenever possible.
%For a complete list of OCIS codes, visit: https://www.osapublishing.org/oe/submit/ocis/

%%%%%%%%%%%%%%%%%%%%%%% References %%%%%%%%%%%%%%%%%%%%%%%%%

%%%%%%%%%%%%%%%%%%%%%%%%%%  body  %%%%%%%%%%%%%%%%%%%%%%%%%%
\section{Introduction}

Subwavelength gratings are simple periodic optical structures that can serve as various functional components in optical systems, such as optical filters~\cite{filter-Magnusson.1992, Chen:17, Hsu:16}, beam splitters~\cite{SMM-BS-feng.2008, SMM-BS-Wang.2008,Wang:17}, and high-efficiency waveguide couplers~\cite{coupler-Vivien:06, Zhang:15,Chen:17}. Among various numerical methods, rigorous coupled wave analysis (RCWA)~\cite{RCWA1, RCWA2} is one of the most widely used methods to accurately analyze the diffraction properties of subwavelength gratings. Despite its high computational efficiency, RCWA has a disadvantage in that its physical interpretation is unclear, which makes it less helpful for designing functional gratings. The modal method, an alternative to RCWA proposed by Botten \textit{et al.}~\cite{modal-method-botten}, describes the physical process of grating diffraction as propagation of grating modes inside the grating and coupling between grating modes and diffraction orders at the interfaces. For low-contrast deeply etched gratings, reflection at the interfaces and the effects of evanescent grating modes can be neglected; therefore, a simplified modal method (SMM) was introduced~\cite{SMM, SMM-Tishchenko.2005}. When the symmetries of the grating modes are considered, the SMM can give approximate analytical expressions for calculating the diffraction efficiencies of these gratings under Littrow incidence. These expressions can be used to solve inverse  grating problems and are very helpful for designing beam splitter gratings~\cite{SMM-BS-feng.2008, SMM-BS-Wang.2008}. Zheng \textit{et al.} also applied the SMM to designing subwavelength gratings under second Bragg incidence~\cite{Zheng.2008b}. The SMM was also developed for application to the design of triangular-groove~\cite{SMM-tri-Zheng.2008} and sinuous-groove gratings~\cite{SMM-sin-Feng.2010}. 

	The SMM has successfully guided the design of low-contrast subwavelength gratings. However, when the index contrast increases, the reflection at the grating interfaces must be considered; therefore, the simple expressions of the diffraction efficiencies no longer work. Karagodsky \textit{et al.} developed a modal method to study the ultra-wideband high reflectivity in high-contrast gratings~\cite{HCG-modal}. On the basis of the method in Ref.~\cite{HCG-modal}, he also explained how the resonance mechanism in high-contrast gratings can be understood in terms of the Fabry--Perot (FP) resonance of the grating modes~\cite{FP-explanation} and proposed a simple mathematical expression to predict the resonance wavelength. Wu \textit{et al.}~\cite{Wu.2013} applied Karagodsky's method to analysis of guided mode resonance (GMR) gratings with asymmetric coatings. However, the method developed by Karagodsky requires even symmetric grating modes, and it is applicable only for normal incidence. In 2013, Sun \textit{et al.} first proposed a multi-reflection model of grating diffraction based on the SMM~\cite{Sun.2013}. Sun's model uses a Fresnel form to express the reflection and transmission coefficients by analogy to those of a flat interface. Yang and Li recently introduced an improved SMM~\cite{Yang.2015} to deal with transverse electric (TE)-polarized incidence at arbitrary angles. In Ref.~\cite{Yang.2015}, Yang and Li discussed the relationship between the Fresnel-form-based multi-reflection model and the scatter matrices obtained by their method, and pointed out that the Fresnel-form-based multi-reflection model neglects the coupling of grating modes and works only when the incident angle is around Littrow incidence. In this letter, we propose a multi-reflection model based on the SMM with a complete matrix description to study the resonance in subwavelength dielectric gratings. This model is applicable to analyzing grating diffraction with incidence at arbitrary angles. A GMR Brewster grating filter is designed, and simulation results obtained using the RCWA and our method are compared to verify the validity of our method. Two resonance conditions based on the multi-reflection model are also proposed for evaluating the resonance wavelength.

\section{Multi-reflection model of grating diffraction}

%figure
\begin{figure}[!htbp]
	\centering
	\includegraphics[width=0.7\linewidth]{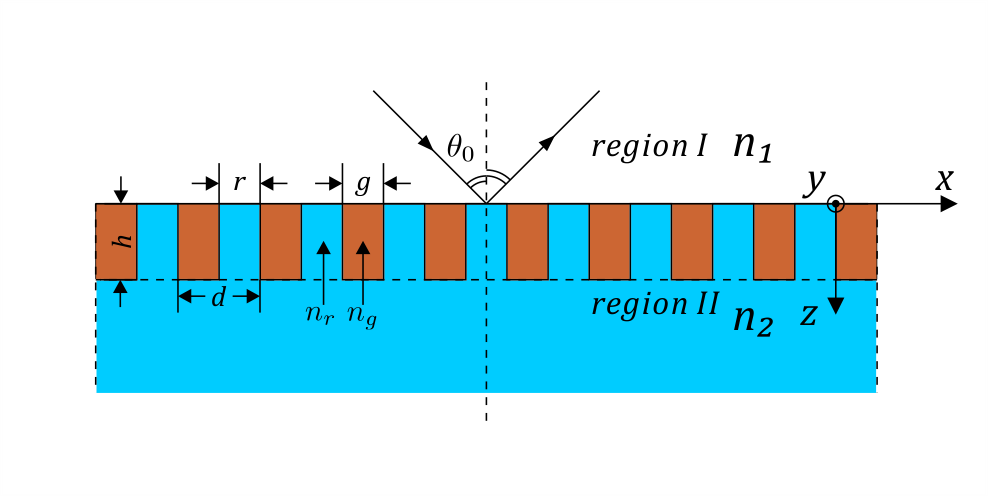}
	\caption{Illustration of a grating and its parameters: the groove depth is $h$, and the grating period is $d$; the widths of the grating ridge and groove are $r$ and $g$, and the corresponding refractive indices are $n_r$ and $n_g$, respectively.}
	\label{fig:grating}
\end{figure}

The grating is schematically illustrated in Fig.~\ref{fig:grating}. The diffraction process in the grating is analogous to the multi-reflection process in a thin film structure: 1.  Part of the incident wave is reflected as diffraction waves by the input plane ($z=0$), and the remaining energy of the incident wave is coupled into downward-propagating modes inside the grating. 2. The downward-propagating modes travel to the output plane ($z=h$). 3. The downward-propagating modes are partially coupled into the transmitted diffraction orders on the output side ($z>h$) and partially reflected as upward-propagating modes. 4. The upward-propagating modes travel to the input plane ($z=0$). 5. The upward-propagating modes are partially coupled into the reflected diffraction orders on the input side ($z<0$) and partially reflected as downward-propagating modes. 6. Steps 2--5 repeat infinitely; the entire transmitted diffraction field is a superposition of all the fields transmitted by the downward-propagating modes in step 3, and the reflected diffraction field is a superposition of the field reflected from the incident wave in step 1 and all the fields transmitted by the up-propagating modes in step 5. The mathematical interpretation of the full process is presented below.

%figure
\begin{figure}[!htbp]
	\centering
	\includegraphics[width=0.5\linewidth]{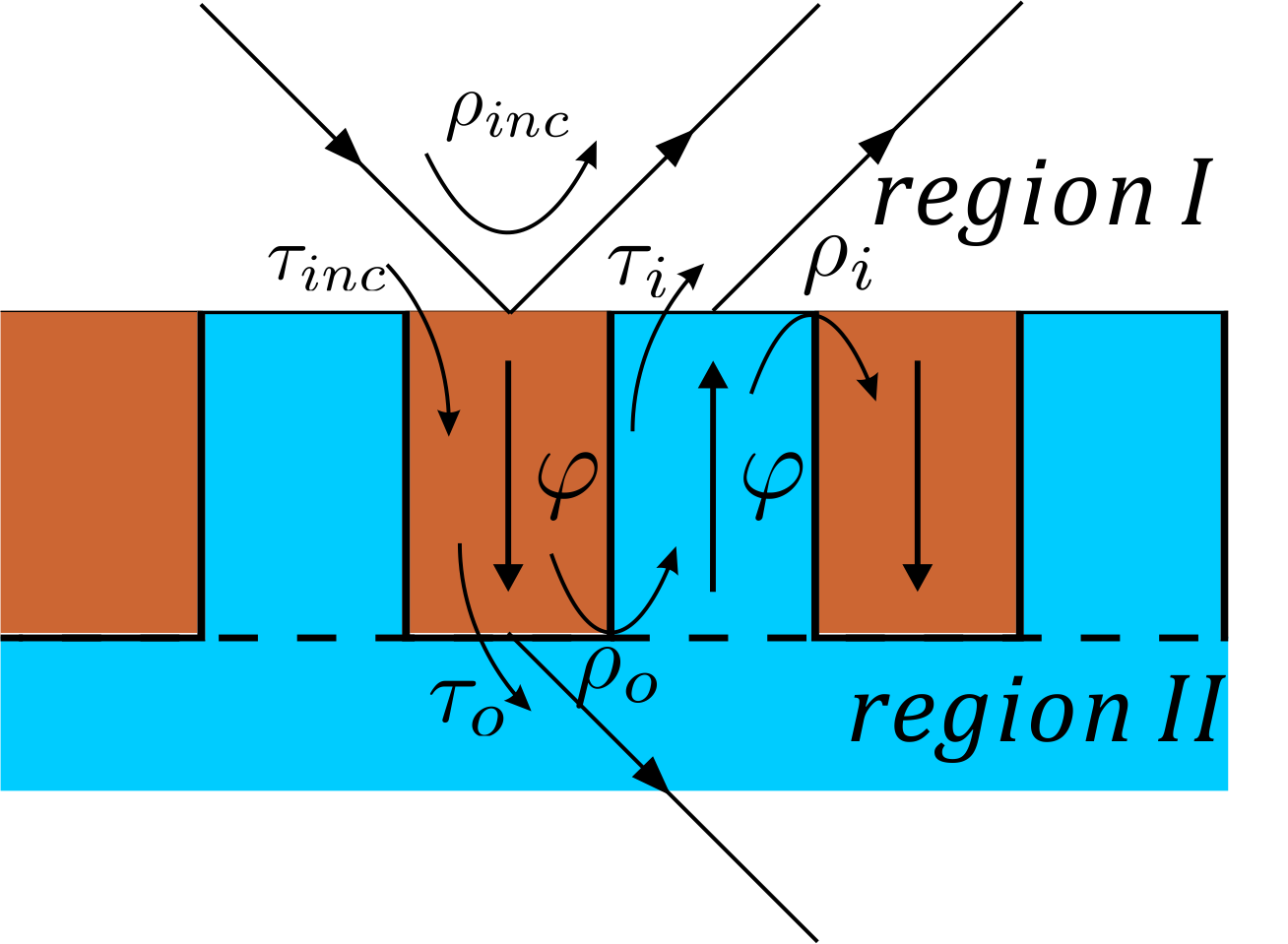}
	\caption{Multi-reflection of grating modes during the diffraction process.}
	\label{fig:multi-refl }
\end{figure}

	In step 1, the field of the incident wave and the diffraction orders reflected from the incident wave by the input interface ($z=0$) at $z<0$ can be written as
\begin{equation}
	\begin{split}
	F(x,z) &= \sum\limits_{n=-\infty}^{+\infty}{\left[  \delta_{n0} \frac{1}{\sqrt{d}} \exp{(\mathrm{i}k_{xn}x)} \exp{(\mathrm{i}k^{I}_{zn}z}) + r_n \frac{1}{\sqrt{d}} \exp{(\mathrm{i}k_{xn}x)} \exp{(-\mathrm{i}k^{I}_{zn}z}) \right]},\\
	F &= \left\{
		\begin{array} {c c}
			E_y	&	\mathrm{TE\ polarization} \\
			H_y	&	\mathrm{TM\ polarization}
		\end{array}  \right.,
	\end{split}
\end{equation}
where $k_{xn} = k_0 \sin{\theta_0} + 2\pi n/d$, $k^{I}_{zn} = \sqrt{(k_0 n_1)^2 - k_{xn}^2}$, and $k_0 = 2 \pi / \lambda$, $r_n$ is the complex amplitude of the $n$th reflection diffraction order reflected from incident wave in this step. The field of transmitted downward-propagating modes inside the grating ($0<z<h$) is~\cite{modal-method-botten}
\begin{equation}
	F(x,z) = \sum\limits_{m=1}^{+\infty}{  a^+_m u_m(x) \exp{(\mathrm{i}\beta_m z}) },
\end{equation}
where $\beta_m=k_0 n^m_{eff}$, $a^+_m$ is the complex amplitude of the $m$th grating mode transmitted from incident wave in this step, and $n^m_{eff}$ is the effective refractive index of the $m$th grating mode which is determined by~\cite{modal-method-botten}
\begin{equation}
	\begin{split}
		\cos{(k_g g)}\cos{(k_r r)} &- \frac{k_g^2 + \gamma^2 k_r^2}{2 \gamma k_g k_r}\sin{(k_g g)}\sin{(k_r r)} = \cos{(k_{x0}d)}, \\
		k_g &= k_0 \sqrt{n_g^2 - (n^m_{eff})^2}, \\
		k_r &= k_0 \sqrt{n_r^2 - (n^m_{eff})^2}, \\
		\gamma &= \left\{
			\begin{array} {c c}
				1	&	\mathrm{TE\ polarization} \\
				n_g^2/n_r^2	&	\mathrm{TM\ polarization}
			\end{array}  \right.,
	\end{split}
\end{equation}
$u_m(x)$ is the unified profile of the $m$th grating mode (detailed form can be found in~\cite{modal-method-botten}) and satisfies
\begin{equation}
	\int\limits_0^d u_m(x) \frac{u^*_n(x)}{\alpha_g(x)} \,\mathbf{d}x = \delta_{mn},
	\label{eq:mode-unif}
\end{equation}
where
\begin{gather*}
	\alpha_g(x) = \left\{
		\begin{array} {l l}
			\mu_r = \mu_g = \mu_0 = 1	&	\mathrm{TE\ polarization} \\
			\epsilon_g(x)	&	\mathrm{TM\ polarization}
		\end{array}  \right., \\
	\epsilon_g(x) = \left\{
		\begin{array} {l l}
			n_g^2	&	id \leq x< r + id \\
			n_r^2	&	r+id \leq x<(i+1)d
		\end{array}  \right. ,
	i = \{\dots, -1, 0, 1, 2, \dots\}.
\end{gather*}

The boundary condition implies that $F(x,z)$ should be matched at $z=0$:
	\begin{equation}
		\sum\limits_{n=-\infty}^{+\infty}{\left[  (\delta_{n0} + r_n) \frac{1}{\sqrt{d}} \exp{(\mathrm{i}k_{xn}x)}  \right]} = \sum\limits_{m=1}^{+\infty}{a^+_m u_m(x)}.
	\end{equation}
If we multiply the left- and right-hand sides by $u^*_m(x)/\alpha_g(x)$, integrate both sides over $\left[0,d\right)$, and use Eq.~\eqref{eq:mode-unif}, we have
	\begin{equation}
		\sum\limits_{n=-\infty}^{+\infty}{J_{mn} (\delta_{n0} + r_n) } = a^+_m, 
		\label{eq:match-h}
	\end{equation}
where 
	\begin{equation}
		J_{mn} = \frac{1}{\sqrt{d}}\int\nolimits_0^d{\exp{(\mathrm{i}k_{xn}x)} \frac{u^*_m(x)}{\alpha_g(x)} \,\mathbf{d}x}.
		\label{eq:j-definition}
	\end{equation}
If we define $\alpha$ as
\begin{gather*}
	\alpha = \left\{
		\begin{array} {l l}
			\mu	&	\mathrm{TE\ polarization} \\
			\epsilon	&	\mathrm{TM\ polarization}
		\end{array}  \right., \\
\end{gather*}
$\partial_z F(x,z)/\alpha(x,z)$ should also be matched at $z=0$:
	\begin{equation}
		\frac{1}{\alpha_I}\sum\limits_{n=-\infty}^{+\infty}{k^I_{zn} (\delta_{n0} - r_n) \frac{1}{\sqrt{d}} \exp{(\mathrm{i}k^I_{xn}x)}} = \sum\limits_{m=1}^{+\infty}{\beta_m a_m^+ \frac{{u_m(x)}}{\alpha_g(x)} }.
	\end{equation}
If we multiply the left- and right-hand sides by $u^*_m(x)$ and integrate both sides over $\left[0,d\right)$, we have
	\begin{equation}
		\frac{1}{\alpha_I}\sum\limits_{n=-\infty}^{+\infty}{L_{mn} k^I_{zn} (\delta_{n0} - r_n)  } = \beta_m a^+_m,
		\label{eq:match-e}
	\end{equation}
where
	\begin{equation}
		L_{mn} = \frac{1}{\sqrt{d}}\int\nolimits_0^d{\exp{(\mathrm{i}k_{xn}x)} u^*_m(x) \,\mathbf{d}x}.
		\label{eq:l-definition}
	\end{equation}
Eqs.~\eqref{eq:match-h}--\eqref{eq:match-e} can be written in matrix form:
\begin{equation}
	\begin{split}
		\mathbf{J} (\mathbf{F} + \mathbf{R}) &= \mathbf{A^+}(0), \\
		\frac{1}{\alpha_I} \mathbf{L} \mathbf{K_I} (\mathbf{F} - \mathbf{R}) &= \mathbf{B} \mathbf{A^+}(0),
	\label{matrix-inc}
	\end{split}
\end{equation}
where $\mathbf{F}, \mathbf{R},  \mathbf{A^+}(0)$ are vectors: $(\mathbf{F})_n = \delta_{n0},  (\mathbf{R})_n = r_n, (\mathbf{A^+}(0))_n = a^+_n \exp{(\mathrm{i} \beta_n z) |_{z=0}}$, and $\mathbf{K_I},  \mathbf{B}$ are diagonal matrices: $(\mathbf{K_I})_{mn} = \delta_{mn} k_{zn}^I$, $\mathbf{B} = \delta_{mn} \beta_n$.
The reflection and transmission matrices for incidence at $z=0$ $\boldsymbol{\rho}_{inc}$ and $\boldsymbol{\tau}_{inc}$ are defined as
\begin{equation}
	\begin{split}
		\mathbf{R} &= \boldsymbol{\rho}_{inc} \mathbf{F}, \\
		\mathbf{A^+}  &= \boldsymbol{\tau}_{inc} \mathbf{F}.
	\end{split}
\end{equation}
Using the mutually inverse relation between $\mathbf{J}$ and $\mathbf{L}$ (proved in the \textbf{Appendix}), 
\begin{equation}
	\begin{split}
	\mathbf{J}^{\mathrm{H}} \mathbf{L} &= \mathbf{I}, \\
	\mathbf{L}^{\mathrm{H}} \mathbf{J} &= \mathbf{I},
	\end{split}
\end{equation}
we can solve $\boldsymbol{\rho}_{inc}$ and $\boldsymbol{\tau}_{inc}$:
\begin{equation}
	\begin{split}
		\boldsymbol{\rho}_{inc} &= \mathbf{L}^{\mathrm{H}} [2 (\alpha_I \mathbf{J}\mathbf{K_I^{-1}}\mathbf{J}^{\mathrm{H}}\mathbf{B} + \mathbf{I})^{-1} - \mathbf{I}] \mathbf{J}, \\
		\boldsymbol{\tau}_{inc} &= 2 (\alpha_I \mathbf{J}\mathbf{K_I^{-1}}\mathbf{J}^{\mathrm{H}}\mathbf{B} + \mathbf{I})^{-1} \mathbf{J}.
	\end{split}
	\label{eq:step1}
\end{equation}
 
	In step 2, downward-propagating modes travel from the input plane ($z=0$) to the output plane ($z=h$), and the weight of each mode changes with phase accumulation: $[a^+_m \exp{(\mathrm{i}\beta_m z)}] |_{z=h} = \exp{(\mathrm{i}\beta_m h)} [a^+_m \exp{(\mathrm{i}\beta_m z)}] |_{z=0}$. The matrix expression of this mode propagation process is
\begin{equation}
	\mathbf{A^+}(h) = \boldsymbol{\varphi} \mathbf{A^+}(0),
	\label{step2}
\end{equation}
where $\boldsymbol{\varphi}$ is a diagonal matrix: $(\boldsymbol{\varphi})_{mn} = \delta_{mn} \exp{(\mathrm{i}\beta_m h)} $.

	Next, we examine step 3 at the input plane, $z=h$. The downward-propagating grating modes travel down from $z=0$, and are reflected as upward-propagating modes by the output plane ($z=h$). The superposed field $F(x,z)$ of these modes in $0<z<h$ is
\begin{equation}
	F(x,z) = \sum\limits_{m=1}^{+\infty}{a^+_m u_m(x) \exp{(\mathrm{i}\beta_m z)} + a^-_m u_m(x) \exp{(-\mathrm{i}\beta_m z})}.
\end{equation}
Here $a^-_m$ is the complex amplitude of the $m$th upward-propagating grating mode reflected from the downward-propagating modes in this step.
The remaining energy of the downward-propagating modes is coupled into the transmitted diffraction waves. The field of diffraction orders transmitted from by the downward-propagating modes in $z>h$ can be written as
\begin{equation}
	F(x,z) = \sum\limits_{n=-\infty}^{+\infty}{ t_n \frac{1}{\sqrt{d}} \exp{(\mathrm{i}k_{xn}x)} \exp{(\mathrm{i}k^{II}_{zn} (z-h))} },
\end{equation}
where $k^{II}_{zn} = \sqrt{(k_0 n_2)^2 - k_{xn}^2}$, $t_n$ is the complex amplitude of the $n$th transmissive diffraction order. As in Eq.\eqref{eq:step1}, the boundary conditions can be written in matrix form:
\begin{equation}
	\begin{split}
		\mathbf{J} \mathbf{R} &= \mathbf{A^+}(h) + \mathbf{A^-}(h), \\
		- \frac{1}{\alpha_I}\mathbf{L}\mathbf{K_{II}} \mathbf{T} &= \mathbf{B}(\mathbf{A^+}(h) - \mathbf{A^-}(h)),
	\end{split}
	\label{eq:matrix-in}
\end{equation}
where $\mathbf{T}, \mathbf{A^+}(h)  and \mathbf{A^-}(h)$ are vectors:  $(\mathbf{T})_n = t_n, (\mathbf{A^+}(h))_n = a^+_n\exp{(\mathrm{i} \beta_n z)|_{z=h}},  (\mathbf{A^-}(h))_n = a^-_n\exp{(-\mathrm{i} \beta_n z) |_{z=h}}$.  $\mathbf{K_{II}}$ is a diagonal matrix: $(\mathbf{K_{II}})_{mn} = \delta_{mn} k_{zn}^{II}$. Denoting the reflection matrix and transmission matrix at the input plane as $\boldsymbol{\rho}_{o}$ and $\boldsymbol{\tau}_{o}$, respectively:
\begin{equation}
	\begin{split}
		\mathbf{A^-(h)} &= \boldsymbol{\rho}_{o} \mathbf{A^+}(h), \\
		\mathbf{T}  &= \boldsymbol{\tau}_{o} \mathbf{A^+}(h),
	\end{split}
\end{equation}
we can solve $\boldsymbol{\rho}_o$, $\boldsymbol{\tau}_{o}$ from Eq.~(\eqref{eq:matrix-in}):
\begin{equation}
	\begin{split}
		\mathbf{\rho}_o &= (\alpha_{II} \mathbf{J}\mathbf{K_{II}^{-1}}\mathbf{J}^{\mathrm{H}}\mathbf{B} + \mathbf{I})^{-1} (\alpha_{II} \mathbf{J}\mathbf{K_{II}^{-1}}\mathbf{J}^{\mathrm{H}}\mathbf{B} - \mathbf{I}),\\
		\boldsymbol{\tau}_o &= \mathbf{L}^{\mathrm{H}}  (\boldsymbol{\rho}_o + \mathrm{I}).
	\end{split}
	\label{eq:step3}
\end{equation}

	Step 4 is similar to step 2. The modes are propagating in the opposite direction to that in step 2:
\begin{equation}
	\mathbf{A^-}(0) = \boldsymbol{\varphi} \mathbf{A^-}(h).
	\label{step4}
\end{equation}

	Using a process similar to that in step 3, by matching $F(x,z)$ and $\partial_z F(x,z)/\alpha(x,z)$ at $z=h$, we can solve the reflection and transmission matrices $\boldsymbol{\rho}_i$ and $\boldsymbol{\tau}_i$ in step 5:
\begin{equation}
	\begin{split}
		\boldsymbol{\rho}_i &= (\alpha_I \mathbf{J}\mathbf{K_I^{-1}}\mathbf{J}^{\mathrm{H}}\mathbf{B} + \mathbf{I})^{-1} (\alpha_I \mathbf{J}\mathbf{K_I^{-1}}\mathbf{J}^{\mathrm{H}}\mathbf{B} - \mathbf{I}), \\
		\boldsymbol{\tau}_i &= \mathbf{L}^{\mathrm{H}}  (\boldsymbol{\rho}_i + \mathrm{I}).
	\end{split}
	\label{eq:step4}
\end{equation}

	After the reflection and transmission matrices are obtained, the total diffraction field can be calculated by superposing all the diffraction fields in the reflection and transmission processes in steps 1--5. The complex amplitude vectors of the total reflected and transmitted diffraction orders are
\begin{equation}
	\begin{split}
		\mathbf{R}_{total}	&= \rho_{inc} \mathbf{F} + \sum\limits_{k=0}^{\infty} {\tau_i \varphi \rho_o \varphi (\rho_i \varphi \rho_o \varphi)^k \tau_{inc} \mathbf{F}} \\
								&= \rho_{inc} \mathbf{F} + \tau_i \varphi \rho_o \varphi \lim_{k \to \infty}  (I - \rho_i \varphi \rho_o \varphi)^{-1}  [I - (\rho_i \varphi \rho_o \varphi)^k] \tau_{inc} \mathbf{F},\\
		\mathbf{T}_{total}	&= \sum\limits_{k=0}^{\infty} {\tau_o \varphi (\rho_i \varphi \rho_o \varphi)^k \tau_{inc} \mathbf{F}} \\
								&= \tau_o \varphi \lim_{k \to \infty}  (I - \rho_i \varphi \rho_o \varphi)^{-1}  [I - (\rho_i \varphi \rho_o \varphi)^k] \tau_{inc} \mathbf{F}.
	\end{split}
\end{equation}
If $\rho_i \varphi \rho_o \varphi$ is a convergent matrix, which it is in most cases, 
\begin{equation}
	\begin{split}
		\mathbf{R}_{total}	&= \rho_{inc} \mathbf{F} + \tau_i \varphi \rho_o \varphi (I - \rho_i \varphi \rho_o \varphi)^{-1} \tau_{inc} \mathbf{F},\\
		\mathbf{T}_{total}	&= \tau_o \varphi (I - \rho_i \varphi \rho_o \varphi)^{-1}  \tau_{inc} \mathbf{F}.
		\label{eq:matrix-eq}
	\end{split}
\end{equation}
Then the diffraction efficiencies can be calculated as
\begin{equation}
	\begin{split}
		\eta^R_{i} = \frac{\mathrm{real}[(\mathbf{K_I})_i] \left| ( \mathbf{R}_{total})_i \right|^2} {k_0 n_1 \cos{\theta_0}},\\
		\eta^T_{i} = \frac{\mathrm{real}[(\mathbf{K_{II}})_i] \left| ( \mathbf{T}_{total})_i \right|^2} {k_0 n_1 \cos{\theta_0}},
	\end{split}
	\label{eq:efficiency}
\end{equation}
where $\eta^R_{i}$ and $\eta^T_{i}$ are the efficiencies of the $i$th reflected and transmitted orders, respectively, and ``real'' indicates the real part of a complex number.

Similar to simplified modal method~\cite{SMM, SMM-Tishchenko.2005}, a propagating mode approximation can be introduced in our model that the effects of evanescent modes are neglected and the modal series expansions  are cut off after all the propagating modes. In this way, the matrices with infinite sizes in Eqs.~\eqref{eq:efficiency}  become $N \times N$ matrices, where N is the number of propagating modes.

\section{Application for GMR Brewster filter}
%figure    
\begin{figure}[!ht]
	\centering
	\includegraphics[width=0.6\linewidth]{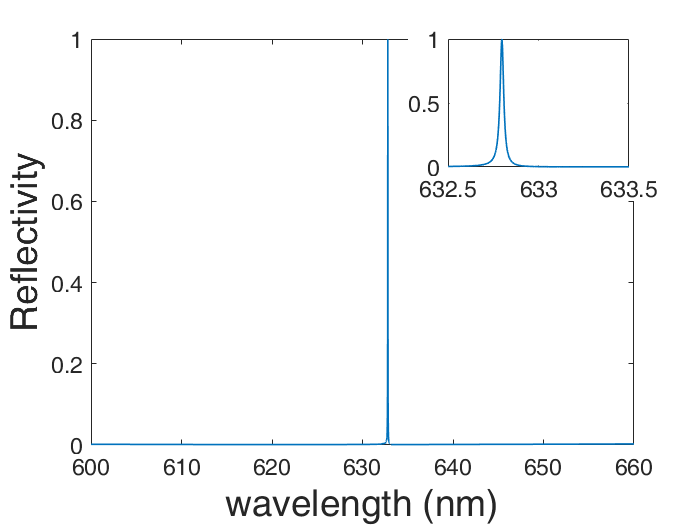}
	\caption{Reflectivity spectrum of the GMR Brewster filter with the parameters $f=0.5$, $d=270.9$\,nm, $h=525.9$\,nm, $n_1=1$, $n_g=2$, and $n_r=n_2=1.457$. The incident angle is $\theta_B=60.7\degree$. Resonance occurs at 632.8\,nm, where there is a sharp transition from 0\% to 100\%.}
	\label{fig:rcwa}
\end{figure}

	The guided mode resonant (GMR) grating~\cite{GMR-grating-Magnusson.1992, GMR-grating-Wang.1993} is a type of high-contrast grating. Most GMR gratings are used under normal incidence; therefore, Karagodsky's method is applicable. The GMR Brewster grating~\cite{GMR-Brewster-Magnusson.1998}, which exploits the extinguished transverse magnetic (TM)-polarized reflection due to the Brewster effect, can be used to implement extremely low-reflectivity sidebands. As no symmetric grating modes exist under Brewster incidence, Karagodsky's method no longer works. 

%figure    
\begin{figure}[!hb]
	\centering
	\includegraphics[width=0.6\linewidth]{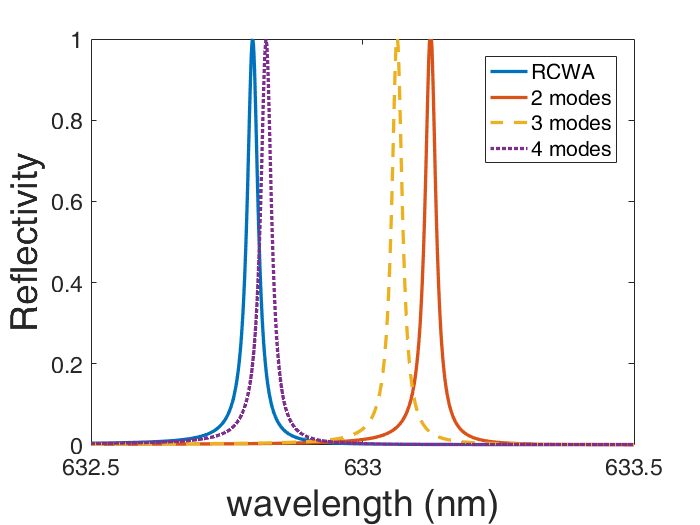}
	\caption{Comparison of simulation results obtained using RCWA and the multi-reflection modal method. When the first four modes are considered, the modal solution is in good agreement with that of RCWA.}
	\label{fig:refl-rcwa-modal}
\end{figure}

	To verify our multi-reflection model, we designed a GMR Brewster grating filter for an incident wavelength of 632.8\,nm. The filter consists of a waveguide grating layer on a fused-silica substrate. The parameters of the filter are $f=0.5$, $d=270.9$\,nm, $h=525.9$\,nm, $n_1=1$, $n_g=2$, and $n_r=n_2=1.457$. The incident wave illuminates the filter at the Brewster angle of this structure, $\theta_B=60.7\degree$, which is calculated by employing thin film theory~\cite{Macleod.2010b}. Fig.~\ref{fig:rcwa} shows the variation of the reflectivity calculated by RCWA. Resonance occurs at a wavelength of 632.8\,nm with a sharp peak. It is worth mentioning that the sharp peak is also a result of co-existence of GMR and Rayleigh anomaly~\cite{GMR-Rayleigh-1, GMR-Rayleigh-2}, as the Rayleigh anomaly condition $k_{x,-1} = k_0\,n_2$ is meeted, where $k_{x,-1}$ is the x-component of the wave vector of the -1st order.

	The first two grating modes inside the waveguide grating are propagating modes. Assuming that these two modes  will dominate the process of diffraction, we introduced the propagating mode approximation and cut off the modal series expansions in Section 2 after the first two modes. The variation of the reflectivity calculated by our method and that calculated by RCWA are compared in Fig.~\ref{fig:refl-rcwa-modal}. Under the two-mode cutoff approximation, the resonance wavelength is located at $\lambda = 633.13$\, nm. When more modes are considered, the peak of the variation moves closer to 632.8\, nm, and the fourth mode has a more significant effect than the third mode on the peak shift. 

%figure
\begin{figure}[!htpb]
	\centering
	\subfloat[$|H_1/H_i|^2$]{
		\includegraphics[width=0.35\linewidth]{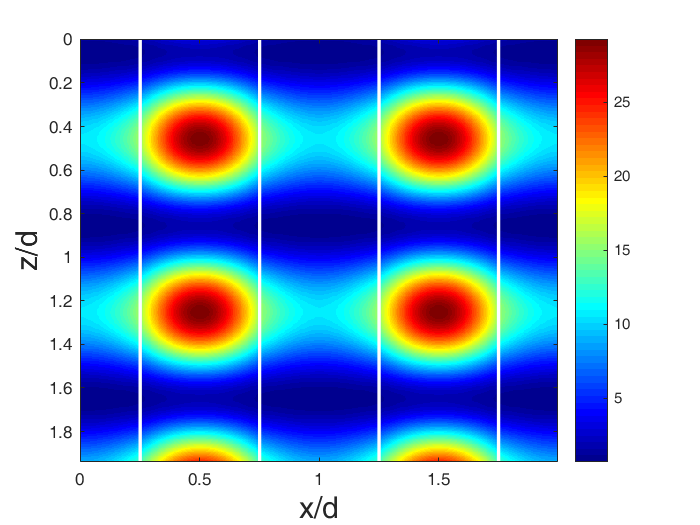}
	}
	\subfloat[$|H_2/H_i|^2$]{
		\includegraphics[width=0.35\linewidth]{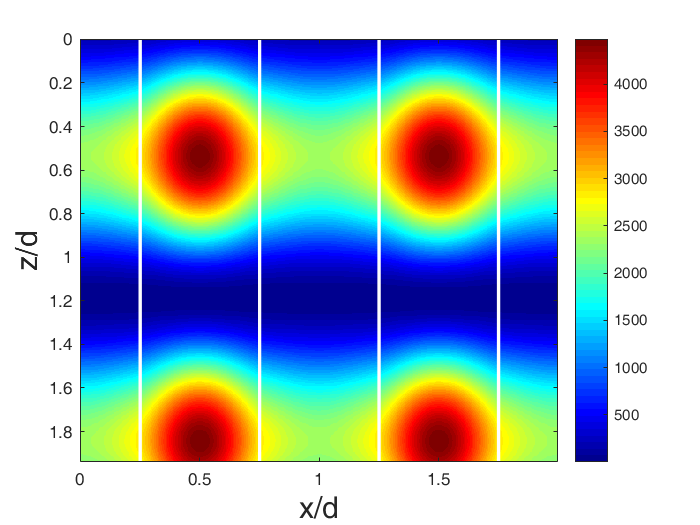}
	}\\
	\subfloat[$|H_3/H_i|^2$]{
		\includegraphics[width=0.35\linewidth]{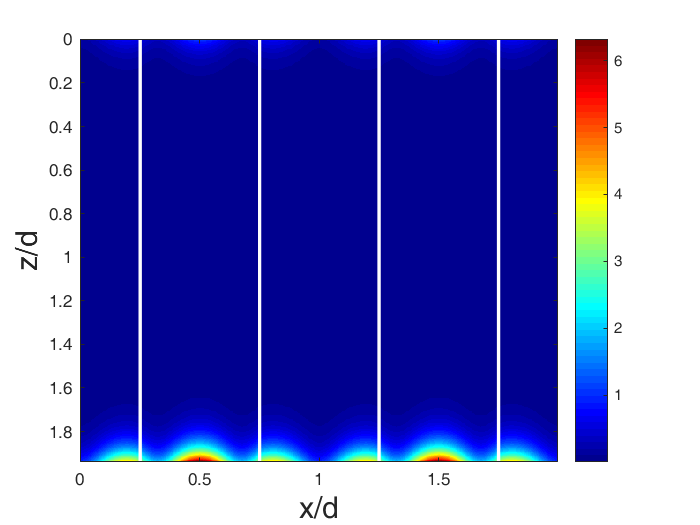}
	}
	\subfloat[$|H_4/H_i|^2$]{
		\includegraphics[width=0.35\linewidth]{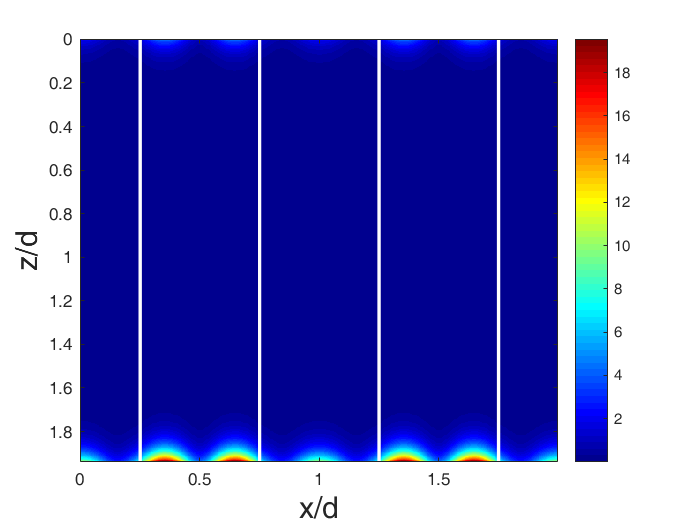}
	}
	\captionsetup{justification=raggedright}
	\caption{Density plot of magnetic field intensity distributions of the first four modes.}
	\label{fig:field-mode}
\end{figure}

	We also plot the intensity distributions of the magnetic fields of the first four grating modes in Fig.~\ref{fig:field-mode}. The first two modes are propagating modes, and they form standing waves in the $z$ direction. The third and fourth modes are evanescent, and most of their energy is confined near the input and output planes ($z=0,h$). The field intensities of the first two modes are much larger than those of the other modes, and the field intensity of the fourth mode is larger than that of the third mode. These results validate our assumption that the effects of the first two modes are dominant, and confirm that the fourth mode has a larger effect than the third mode in this grating.

\iffalse
	 By using a uniform medium approximation and the slab waveguide theory, one can evaluate the resonance wavelength by solving the eigenvalue equation~\cite{GMR-grating-Wang.1993}:
\begin{equation}
    \tan(\kappa_i h) = \frac{\epsilon_e \kappa_i (\epsilon_I \gamma_i + \epsilon_{II} \delta_i)}{\epsilon_I \epsilon_3 \kappa_i^2 - \epsilon_e^2 \gamma_i \delta_i},
    \label{eq:reso-uni-slab}
\end{equation}
where $\kappa_i = \sqrt{\epsilon_e k^2 - \beta_{si}^2}$, $\gamma_i = \sqrt{\beta_{si}^2 - \epsilon_1 k^2}$, and $\delta_i = \sqrt{\beta_{si}^2 - \epsilon_2 k^2}$. Further, $\beta_{si}$ is the propagation constant of the $i$th waveguide mode in the slab:
\begin{equation}
    \beta_{si} = k(\sqrt{\epsilon_e} \sin{\theta_B} - i \lambda/d).
    \label{eq:beta-slab}
\end{equation}
$\epsilon_e$ is the effective relative permittivity of the waveguide grating according to the effective medium theory:
\begin{equation}
 	\epsilon_e = [f n_c^{-2} + (1-f) n_g^{-2}]^{-1}.
	\label{eq:eff-medium}
\end{equation}
In this way, the resonance wavelength is evaluated as 630.95\,nm.
\fi

	The multi-reflection model can also give the matrix FP resonance condition to predict the resonance wavelength: resonance occurs when an assembly of modes $\mathbf{M}$ in the grating layer can constructively interfere with themselves after a full round trip according to steps 2--5. This condition can be described by the expression $\mathbf{M}= \boldsymbol{\rho}_i \boldsymbol{\varphi \rho}_o\boldsymbol{\varphi M}$, which is equivalent to 
\begin{equation}	
	\mathrm{det}(\mathbf{I} - \boldsymbol{\rho}_i \boldsymbol{\varphi \rho}_o\boldsymbol{\varphi})=0 .
	\label{eq:det-condition}
\end{equation}		 

%figure    
\begin{figure}[!htbp]
	\centering
	\includegraphics[width=0.6\linewidth]{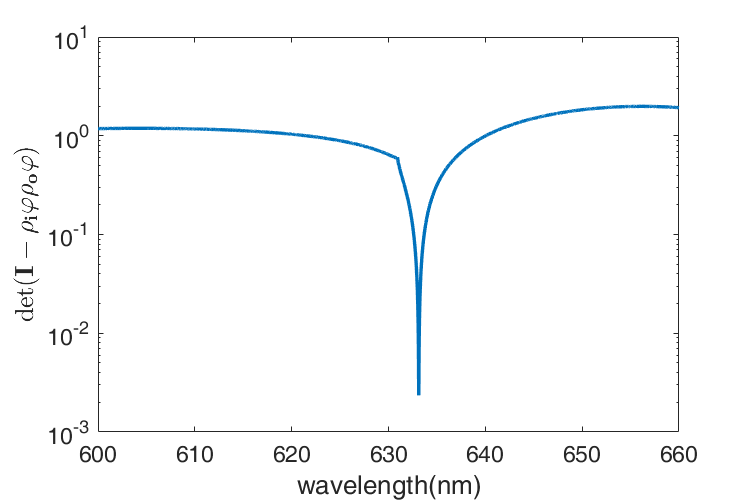}
	\caption{Value of $\mathrm{det}(\mathbf{I} - \boldsymbol{\rho}_i \boldsymbol{\varphi \rho}_o\boldsymbol{\varphi})$. The minimum is located at 633.1\,nm.}
	\label{fig:detvalue}
\end{figure}	

	The values of $\mathrm{det}(\mathbf{I} - \boldsymbol{\rho}_i \boldsymbol{\varphi \rho}_o\boldsymbol{\varphi})$ for different incident wavelengths are calculated and presented in Fig.~\ref{fig:detvalue}. Under the two-mode cutoff approximation, the minimum of this determinant is located at 633.1\,nm, which is close to the resonance wavelength. Compared to the evaluation by the method in Ref.~\cite{GMR-grating-Wang.1993}, the prediction by our method is much closer to the resonance wavelength obtained using RCWA. It is also worth noting that the filter is designed by using this resonance condition and simulated annealing algorithm.

\section{Discussion}
%figure
\begin{figure}[!htbp]
	\centering
	\subfloat[Value of $ 1- \left| \boldsymbol{\rho}_{i,11}  \boldsymbol{\rho}_{o,11} \right|$]{
		\includegraphics[width=0.45\linewidth]{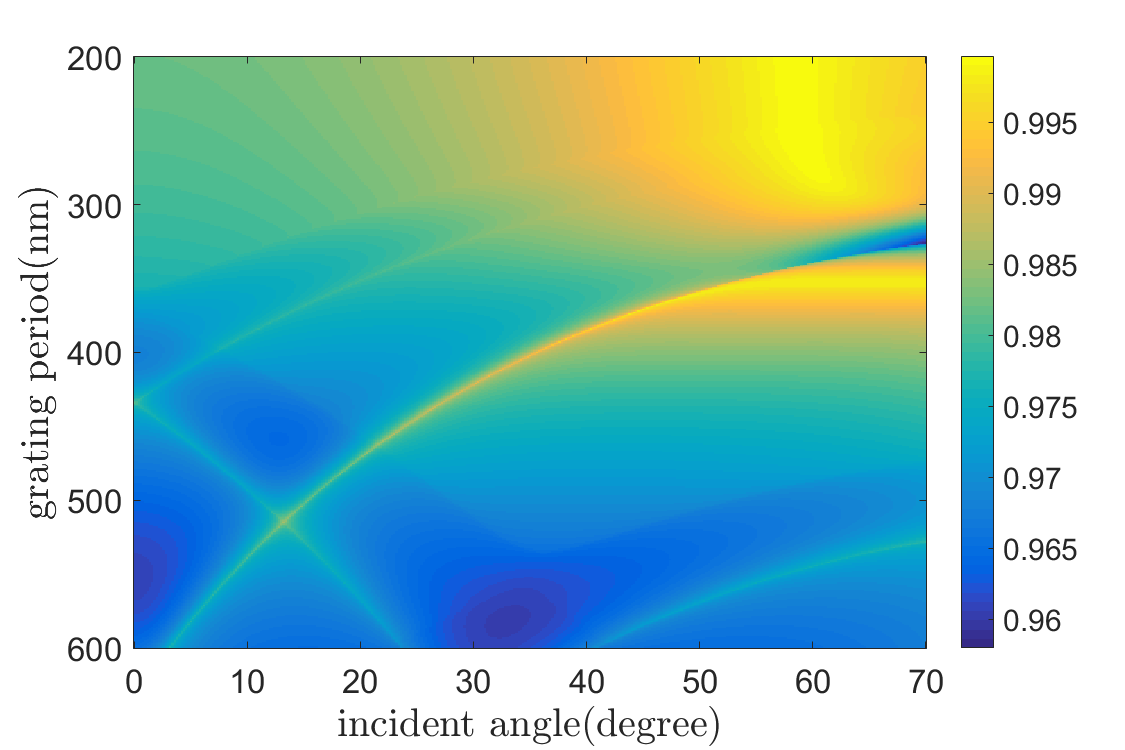}
	}
	\subfloat[Value of $ 1- \left| \boldsymbol{\rho}_{i,22}  \boldsymbol{\rho}_{o,22} \right|$]{
		\includegraphics[width=0.45\linewidth]{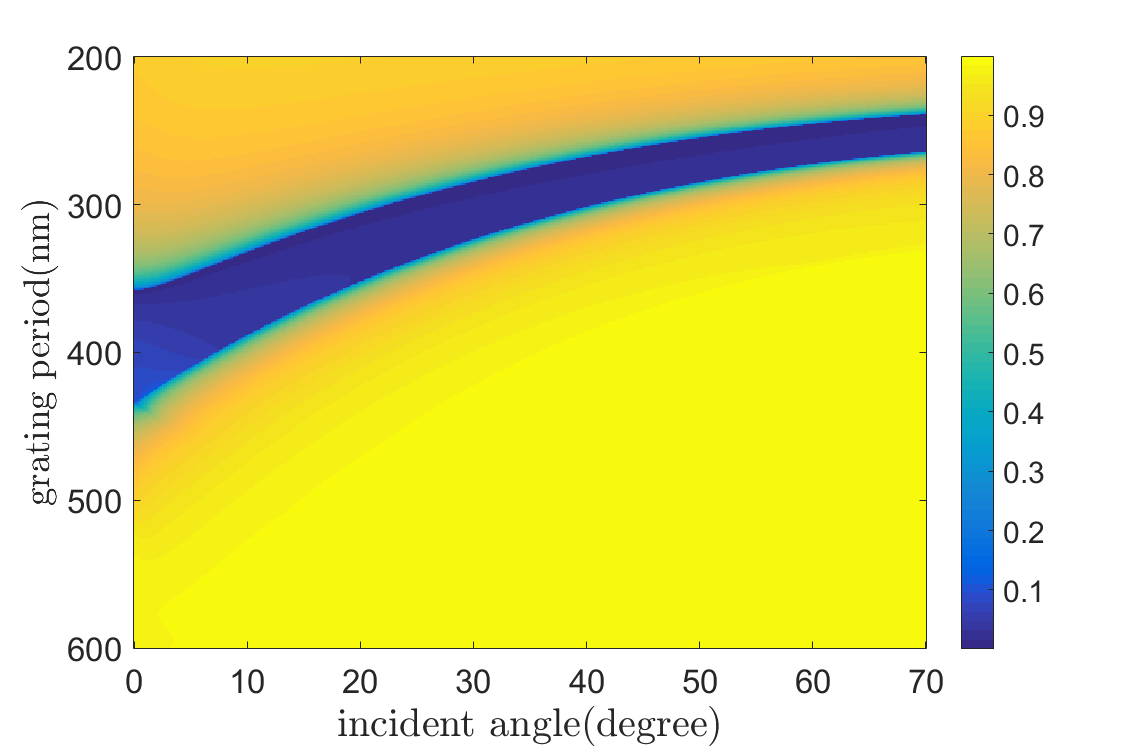}
	}\\

	\subfloat[Value of $ 1- \left| \boldsymbol{\rho}_{i,33}  \boldsymbol{\rho}_{o,33} \right|$]{
		\includegraphics[width=0.45\linewidth]{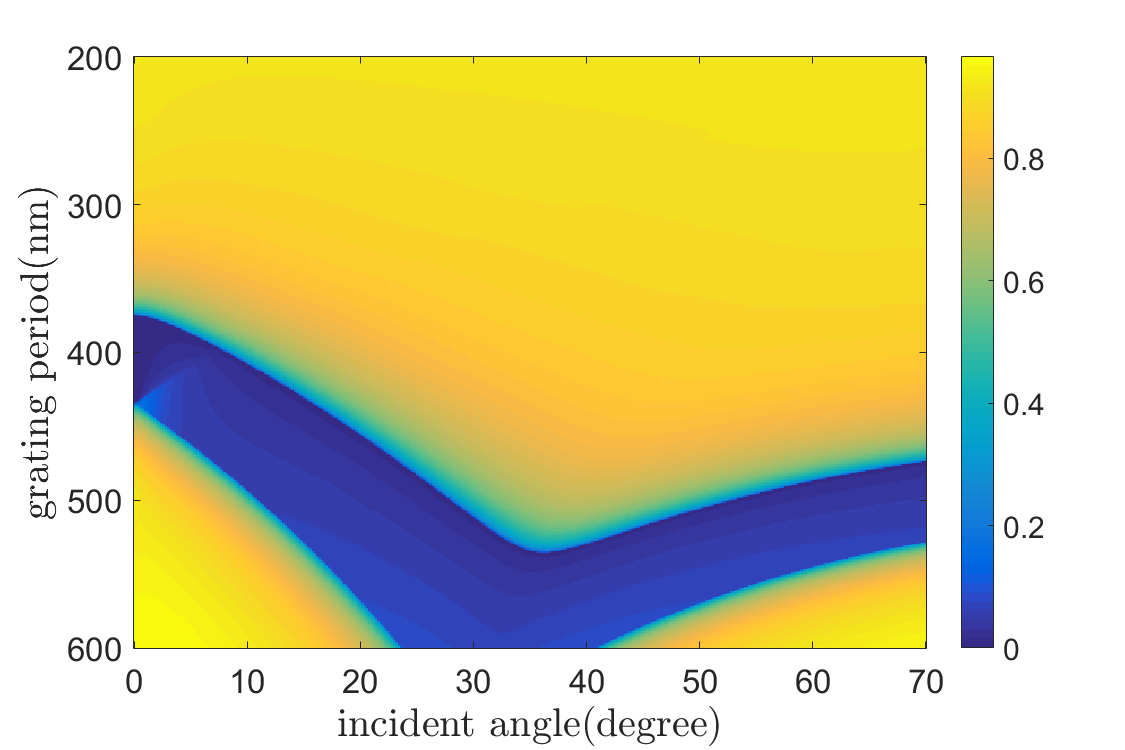}
	}
	\subfloat[Minimum of $\left|\mathrm{det}(\mathbf{I} - \boldsymbol{\rho}_i \boldsymbol{\varphi \rho}_o\boldsymbol{\varphi})\right|$]{
		\includegraphics[width=0.45\linewidth]{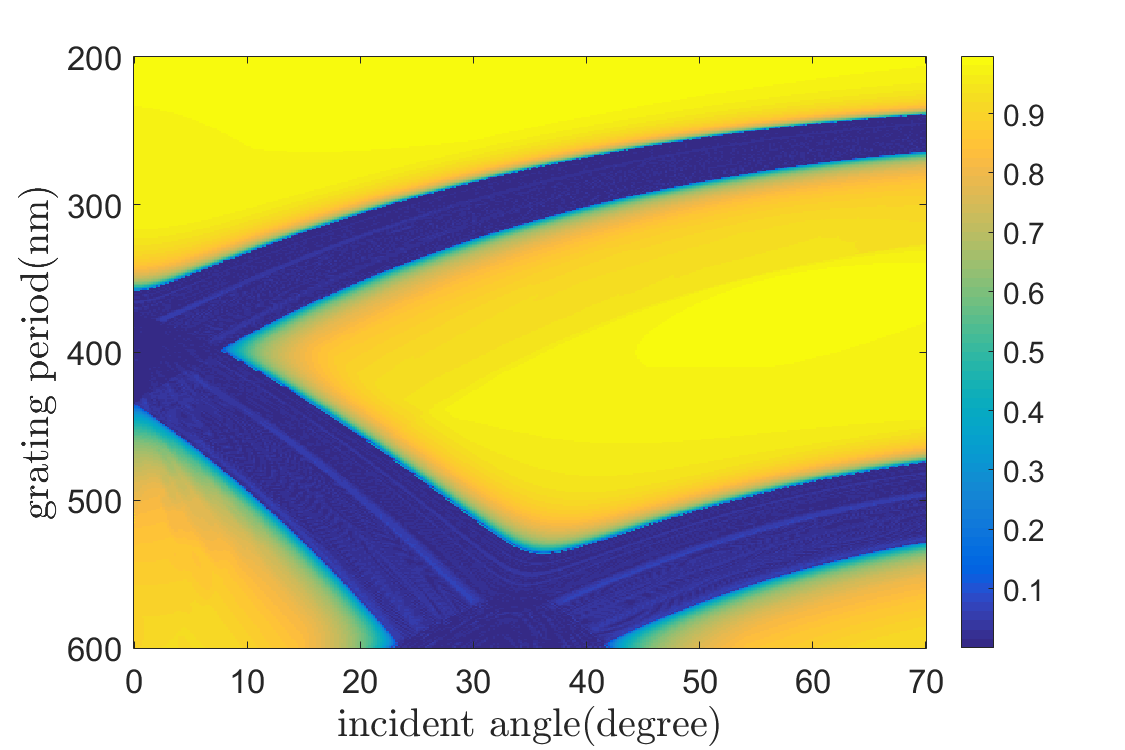}
	}
	\captionsetup{justification=raggedright}
	\caption{Anticipation of resonance by using the conditions Eq.\eqref{eq:one-mode} and Eq.\eqref{eq:det-condition}}
	\label{fig:condition}
\end{figure}

	If one of the propagating modes can constructively interfere with itself after a full round trip in the grating, resonance may also happen. In this case, the condition Eq.\eqref{eq:det-condition} can be simpler as:
\begin{gather}	
	1 - \boldsymbol{\rho}_{i,qq}  \boldsymbol{\rho}_{o,qq}  \exp{(\mathrm{i} 2 \beta_q h)} = 0  \label{eq:one-mode-a} \\
	\Rightarrow \left| \boldsymbol{\rho}_{i,qq}  \boldsymbol{\rho}_{o,qq} \right| = 1. \label{eq:one-mode}
\end{gather}	
where $q$ means the $q$th mode, $\boldsymbol{\rho}_{i,qq}, \boldsymbol{\rho}_{o,qq}$ mean the $q$th diagonal element of the matrices $\boldsymbol{\rho}_i, \boldsymbol{\rho}_o$. Fig.\ref{fig:condition}(a,b) shows the value of $ 1- \left| \boldsymbol{\rho}_{i,qq}  \boldsymbol{\rho}_{o,qq} \right|$ for the first three modes when the corresponding grating parameters are in the domain: $n_1=1$, $n_g=2$, $n_r=n_2=1.457$, $f = 0.5$, $d \in [200,600]$nm, $\lambda = 632.8$nm, $\theta_0 \in [0\degree, 70\degree]$. In the dark area of Fig.\ref{fig:condition}(b,c),  $1- \left| \boldsymbol{\rho}_{i,qq}  \boldsymbol{\rho}_{o,qq} \right|$ is very close to zero, implying resonance may happen  when the grating depth $h$ satisfies Eq.\eqref{eq:one-mode-a}. 
	
	We can also anticipate the occurrence of resonance in the same domain with the condition Eq.\eqref{eq:det-condition} (by checking whether there is an $h$ can satisfy Eq.\eqref{eq:det-condition}),  since its validity has already been proved (Fig.\ref{fig:detvalue}). To find whether such an $h$ exists for a certain point in this domain, we have checked if the minimum of the determinant among $h \in [0,3000]$nm for that point is close to 0, and these minima for all the points in this domain are shown in Fig.\ref{fig:condition}(d) (all the determinants are calculated under 6 modes cutoff approximation). The dark area in Fig.\ref{fig:condition}(d) almost coincides with those in Fig.\ref{fig:condition}(b) and Fig.\ref{fig:condition}(c), indicating that Eq.\eqref{eq:one-mode} successfully anticipates the occurrence of resonance in this domain. However, it is not expected that there isn't any other dark areas in Fig.\ref{fig:condition}(c). It needs a further investigation that whether the first mode and other modes can constructively self interfere and result resonances in other cases.

\section{Conclusion}
In conclusion, we introduced a multi-reflection model based on the simplified modal method (SMM) for analyzing diffraction in subwavelength gratings. Our method is successfully applied to analyze GMR gratings under non-normal incidence. The simulation result for a GMR Brewster filter obtained by using our method is compared with that using RCWA. Even when only two modes are used in this specified cases, the multi-reflection modal solution is in good agreement with the RCWA solution. Our multi-reflection model also provides a matrix FP resonance condition to evaluate the resonance wavelength. A single-mode resonance condition  is also developed from the matrix resonance condition. The expression of the single-mode resonance condition is simple and its clear physics view for GMR inside the grating and may be useful for designing GMR gratings. Simulation results also infer that the second mode may play a special role in the resonance inside the grating. This interesting fact needs further study to confirm and explain.

\section*{Acknowledgments}
The authors acknowledge the support of the Shanghai Science and Technology Committee (15JC1403500, 16DZ2290102) and Chinese Academy of Sciences (QYZDJ-SSW-JSC014).

\begin{appendices}
\section*{Appendix}
	\begin{equation}
		\frac{1}{d}\int\nolimits_0^d{\exp{(\mathrm{i}k_{xn}x)} \exp{(-\mathrm{i}k_{xp}x)} \,\mathbf{d}x} = \delta_{np}
		\label{eq:pw-unif}
	\end{equation}
Eqs.~\eqref{eq:j-definition}--\eqref{eq:l-definition} are equivalent to
	\begin{equation}
		\frac{1}{\sqrt{d}}\exp{(\mathrm{i}k_{xn}x)} = \sum\limits_q{ j_{qn} u_q(x)},
		\label{eq:expansion1}
	\end{equation}

	\begin{equation}
		\frac{1}{\sqrt{d}}\exp{(-\mathrm{i}k_{xp}x)} = \sum\limits_m{ l^*_{mp} \frac{u^*_m(x)}{\epsilon_g(x)}}.
		\label{eq:expansion2}
	\end{equation}

Substituting Eqs.~\eqref{eq:expansion1} and \eqref{eq:expansion2} into the left side of Eq.~\eqref{eq:pw-unif} yields
	\begin{equation}
	\begin{split}
		\int\nolimits_0^d{\sum\limits_q{ j_{qn} u_q(x)} \sum\limits_m{ l^*_{mp} \frac{u^*_m(x)}{\epsilon_g(x)}} \,\mathbf{d}x} &= \sum\limits_q \sum\limits_m j_{qn} l^*_{mp} \int\nolimits_0^d u_q(x) \frac{u^*_m(x)}{\epsilon_g(x)} \,\mathbf{d}x \\ 
		&= \sum\limits_q \sum\limits_m j_{qn} l^*_{mp} \delta_{mq} \\
		&= \sum\limits_m l^*_{mp} j_{mn} .
	\end{split}
	\end{equation}

And this is equal to the right side:
	\begin{equation}
		\sum\limits_m l^*_{mp} j_{mn} = \delta_{np} .
	\end{equation}

The matrix form is
	\begin{align}
		\boldsymbol{L^H}\boldsymbol{J} &= \boldsymbol{I}, \\
		\Leftrightarrow\ \boldsymbol{J^H}\boldsymbol{L} &= \boldsymbol{I}.
	\end{align}

Similarly,
	\begin{align}
		\int\nolimits_0^d{u_m(x) \frac{{u^*_n(x)}}{\epsilon_g(x)}\,\mathbf{d}x} &= \delta_{mn} \nonumber \\
		\Rightarrow\ \boldsymbol{L}\boldsymbol{J^H} &= \boldsymbol{I} \\
		\Leftrightarrow\ \boldsymbol{J}\boldsymbol{L^H} &= \boldsymbol{I}.
	\end{align}
\end{appendices}

\end{document}